# Colour-Field Based Particle Categorization for Residual Stress Detection and Reduction in Solid SPH Simulations

Gizem Kayar[1][a]
[1]Computer Science Department, New York University, 251 Mercer Street, New York, United States of America

Abstract: Residual stress remains in an object even in the absence of external forces or thermal pressure, which, in turn, may cause significant plastic deformations. In case the residual stress creates unwanted effects on the material and so is undesirable, an efficient solution is necessary to track and eliminate this stress. Smoothed Particle Hydrodynamics has been extensively used in solid mechanics simulations and the inherent colour-field generation approach is a promising tracker for the residual stress. In this paper, we propose a way to use the colour-field approach for eliminating the residual stress and prevent the undesirable premature failure of solid objects.

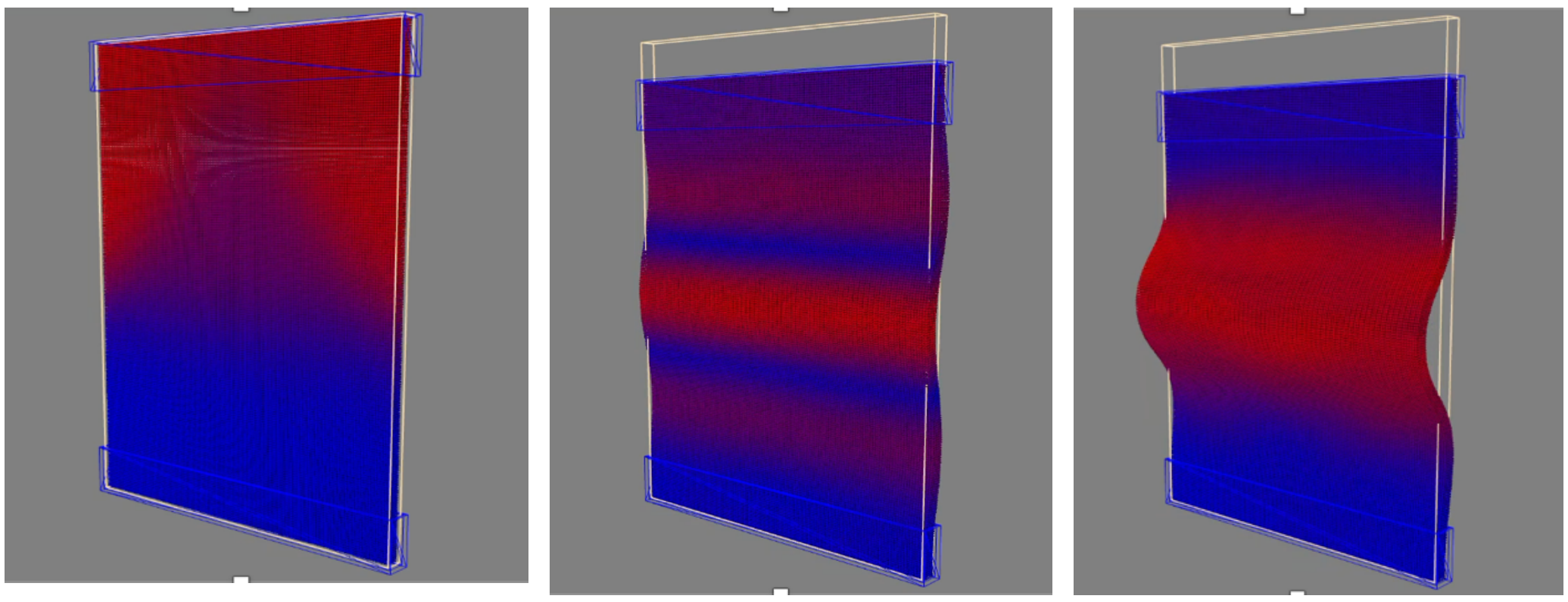

Figure 1. Compressed ductile material implemented with our SPH model and visualized with pressure-based colouring.

## 1 INTRODUCTION

Residual stresses describe any stress remain in the material even after the external forces and thermal effects are excluded from the environment. These stresses may originate from many causes, e.g. cooling rates, volume changes, etc. and they may be created by welding, rolling, forging, casting, machining, heat treatments or surface treatments. After that point, residual stress is generated when an object is stressed beyond its elastic limit, resulting in plastic deformation.

Residual stress should be managed properly for the design, manufacturing, and maintenance phases. Although it may be desirable in some certain conditions, residual stress generally causes the material to fail prematurely. It may affect the fatigue life, stability, resistance and also brittle fractures of objects. Elimination process of residual stresses are extensively researched in material science. Simulations of such conditions help the researchers to understand the problem further and therefore, to create a solution.

[a] https://orcid.org/0000-0002-7811-9357

Since its introduction (Gingold & Monaghan, 1977; Lucy, 1977), Smoothed Particle Hydrodynamics (SPH) has been accepted as one of the major development mediums for the mechanics of continuum media. Although it was introduced for astrophysical problems, later it has been used in various research fields including but not limited to fluid simulations (e.g. Solenthaler & Pajarola, 2009; Shadloo et al., 2016; Gissler et al., 2019) and solid mechanics (e.g. Libersky & Petschek, 1990; Libersky et al., 1993). Although SPH is still in the process of development, the method has been drastically improved over the years to overcome some major, inherent problems. SPH is still one of the easy-to-use numerical methods to model complex systems.

Residual stress has been generated and/or determined by researchers in some SPH simulations for different purposes, such as arc welding (Das & Cleary, 2010), friction stir welding (Eivani et. al, 2021), etc. In our experiments, we observed that the undesired residual stress is the major cause of instability in SPH solid simulations.

Our contribution in this ongoing work is to eliminate the undesired residual stress behaviour from solid mechanics simulations by applying a colour-field technique which is extensively used in SPH fluid simulations. Colour-field approach helps us to identify the particles with potential residual stress which, in turn, are handled differently to reduce the overall stress.

## 2 RELATED WORK

SPH has started to be researched in computer graphics for the last couple of years and still extensively researched in material science. For instance, Ihmsen et al. discusses the force computation problems and visual inefficiency in granular SPH in (Ihmsen et al, 2013). Here, they propose to use a coarsely sampled simulation for force computations, and couple high resolution set of particles to base particles via up-sampling which improves both the computation time and visual quality. In (Nguyen et al, 2017), Nguyen et al. discuss how to use viscous damping and stress regularisation for granular flows. Later, Ghaitanellis et al. proposed a new elastic-viscoplastic model for granular flows in (Ghaitanellis et al, 2018).

In 2016, Pan et al. proposed an application where SPH is used create solitary wave impact on offshore platforms to understand the interaction between large waves and offshore structures which is a good example of SPH's usage in civil engineering and geo-numerics field (Pan et al., 2016). In the same year, Holmes et al. proposed a model for grain-scale fluid flows in porous rocks (Holmes et al., 2016).

Regarding our focus, the elimination process of residual stresses is vital especially when the stress is unwanted. Simulations of such conditions help the researchers to understand the problem further and therefore, to create better solutions.

Residual stress has been generated and/or determined by researchers in some SPH simulations for different purposes, such as arc welding (Das & Cleary, 2010), friction stir welding (Eivani et. al, 2021), etc. Eivani shows a specific example on AZ91 Mg alloy where they combine SPH with neuro-fuzzy computations and ultrasonic testing.

Another specific example comes from Saleh, Luzin and Spencer where they analyse the residual stress in cold spray technique using some numerical and empirical methods (Saleh et al, 2014). Liu and his fellow teammates also propose another technique for the numerical simulation and elimination of residual stress by using shot peening in (Liu et al., 2019)

Discussing other fields, such as fluids, is beyond the scope of this paper. However, state of the art reports like (Koschier et al, 2022) are some of the good resources for those who are interested in the topic.

## 3 SPH

SPH is a method of discretizing spatial quantities using a set of particles equipped with a kernel function. Each particle is defined by a position, mass and a support radius where mass can be computed as:

$$m = \rho \cdot V \quad (1)$$

with $\rho$: fluid density and $V$: particle volume. The word "smoothed" in SPH comes from the smoothing operation which actually means calculating any physical quantity of the particle using the weighted sum of the same quantity of the neighboring particles that lie in the range of a kernel function. So, after the continuous approximation is discretized, he smoothing operation looks as:

$$A_a = \sum_b m_b \frac{A_b}{\rho_b} W(\mathbf{x}_a - \mathbf{x}_b, h) \quad (2)$$

where $A$ is an arbitrary scalar quantity, $\mathbf{x}$ is the position, $b$ is the iterator over all contributing particles and $h$ is the smoothing length.

As we stated before, contributions of the neighbouring particles are governed by the kernel function and can be calculated as, e.g.:

$$W(\mathbf{x}_a - \mathbf{x}_b, h) = \frac{\sigma}{h^d} \begin{cases} 6q^3 - 6q^2 + 1 & 0 \le q \le \frac{1}{2} \\ 2(1-q)^3 & \frac{1}{2} < q \le 1 \\ 0 & q > 1 \end{cases}$$

$$q = |\mathbf{x}_a - \mathbf{x}_b|/h \quad (3)$$

$\sigma$ varies depending on the dimensionality of the system.

In such a system, for both fluids and solids, equations of motion in Lagrangian perspective can be given as:

$$m_i \frac{\partial \mathbf{v}_i}{\partial t} = -V_i \nabla p_i + V_i \mu \nabla^2 \mathbf{v}_i + V_i \mathbf{f}_i \quad (4)$$

which, in turn, can be written as the sum of forces as:

$$m_i \frac{\partial \mathbf{v}_i}{\partial t} = \mathbf{F}_\mathbf{i} \quad (5)$$

$$\mathbf{F}_\mathbf{i} = \mathbf{F}_i^{pressure} + \mathbf{F}_i^{viscosity} + \mathbf{F}_i^{external} \quad (6)$$

Here, forces can be computed using SPH interpolation and later be integrated using one of the explicit or implicit numerical integration schemes, e.g. Euler-Cromer, Verlet, etc, With this approach, thousands to billions of particles can be simulated efficiently (see Figure 2).

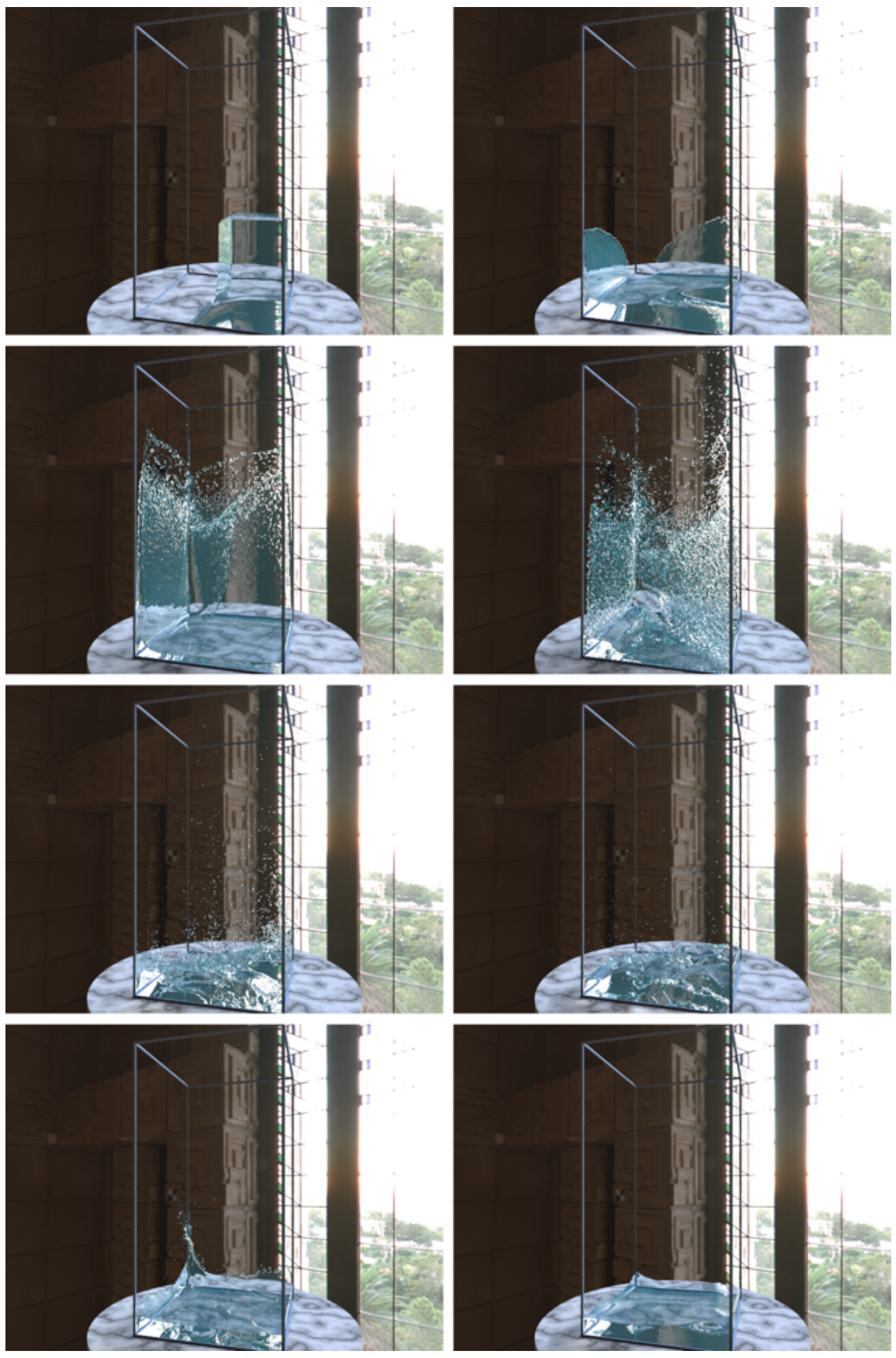

Figure 2: Our SPH fluid simulation with 1.7 million particles.

## 4 SPH FOR SOLIDS AND RESIDUAL STRESS DETECTION

It is important to mention that all solid SPH simulations in our work rely on the methodology discussed in the previous section.

However, the SPH method should be extended so that it may reflect the solid behaviour and yielding criterion (see Figure 1 and Figure 3). Therefore, we firstly implemented elastoplastic solid behaviour in our system by integrating the momentum equation:

$$\frac{\mathrm{d}v^i}{\mathrm{d}t} = \frac{1}{\rho} \frac{\partial \sigma^{ij}}{\partial x^j} + g^i \quad (7)$$

where

$$\sigma^{ij} = -P\delta^{ij} + S^{ij} \quad (8)$$

with pressure $P$, deviatoric stress tensor $S$. Stress can be computed using Hooke's Law.

At this stage, maximum distortion energy criterion can be introduced to the system to estimate the yield of ductile materials. We can even compare the material's yield stress to Von Misses stress to observe the its resistance and yielding thresholds.

Our main focus in this ongoing work is to determine particles which are potentially carrying residual stress. We therefore propose to use the colour field approach. This approach was first proposed by Müller et al. in (Müller et. al, 2003) for determining surface particles in fluid flows. The same approach helps us to categorize the solid particles. We can identify a particle as a surface particle in two conditions combined: 1) if it has less than a certain number of neighbouring particles and 2) if its surface normal **n** is showing more than a meaningful threshold. Surface normal in this situation can be computed as:

$$\mathbf{n} = \nabla c_s \qquad (9)$$

and

$$c_s(\mathbf{p}_i) = \sum_b \frac{m_b}{\rho_b} W(\mathbf{p}_i - \mathbf{p}_j, h) \qquad (10)$$

After identifying surface particles, we generated one more additional layer right behind the surface particles and we constrain those two sets to apply only hydrostatic stress to other particles.

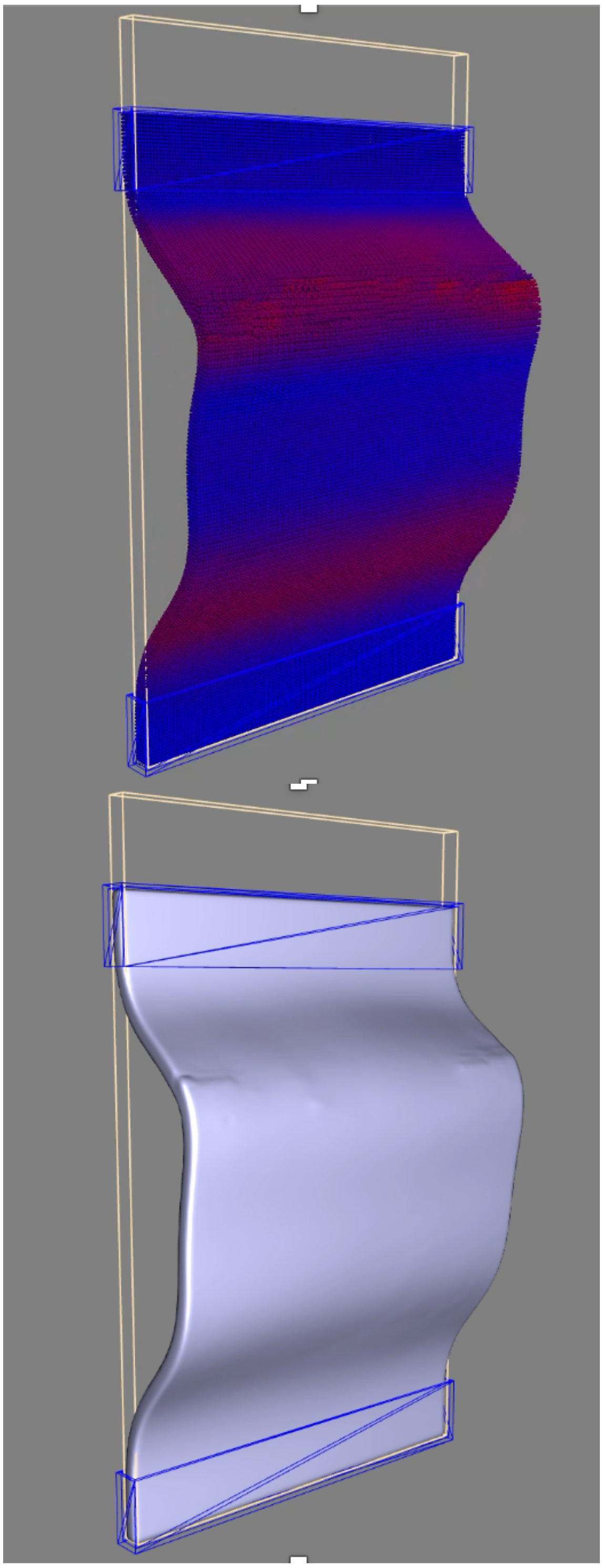

Figure 3. Compressed iron with pressure based particle coloring and marching cubes based surface reconstruction.

## 5 RESULTS

During this approach, we observed that the stability of the system improved significantly and we could use larger time steps for our simulations. To be specific, our simulation time steps has been increased almost twice for all presented scenes. Additionally, we could prevent all undesired crashes and undesired forces from our system.

As an unexpected effect, we also noticed that we could simulate brittle fractures without using an ad-hoc damage model (see Figure 4).

## 6 CONCLUSIONS

In this paper, we presented a technique to improve the stability of a solid system simulation by using smoothed particle hydrodynamics. Based on the conventional SPH model, we integrated elastoplastic solid physics and used the idea of surface extraction for isolating the force computation on certain fields. We observed that the technique works well for both ductile and brittle materials.

This is obviously an ongoing work and needs further improvement. We plan to utilize more test scenes and various comparisons for different material properties in the future. Furthermore, we only discussed the general stability but we did not discuss the performance in detail. In the future, we would like to investigate parallelization techniques for the model.

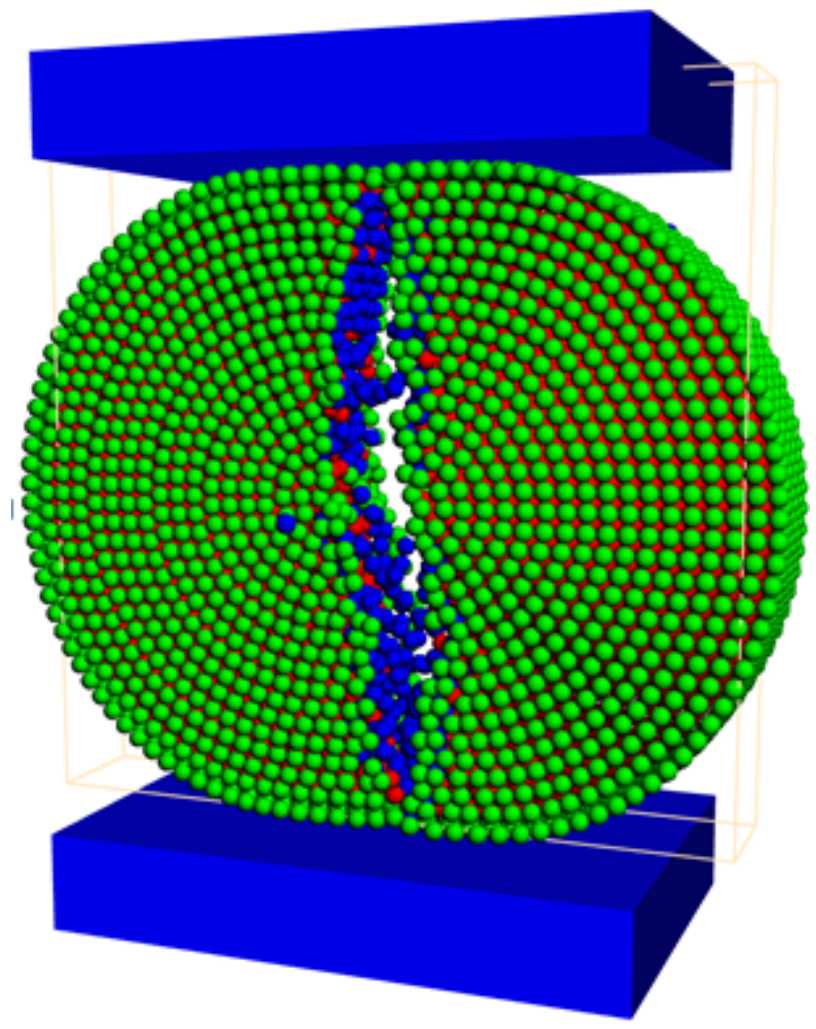

Figure 4. Surface particle generation (green), inner layer additional surface set particles (red) and completely fully particles (blue) during a brittle fracture.